\documentclass[aps,twocolumn,showpacs,groupedaddress]{revtex4}
\usepackage{amssymb}
\usepackage{graphicx}
\usepackage{amsmath}

\newcommand{\ket}[1]{|#1\rangle}
\newcommand{\bra}[1]{\langle#1|}

\begin{document}

\title{An alternative non-Markovianity measure by  divisibility of  dynamical map}
\author{S. C. Hou$^{1}$, X. X. Yi$^{1,2}$, S. X. Yu$^{2,3}$ and C. H. Oh$^{2}$}
\affiliation{$^1$School of Physics and Optoelectronic Technology,
Dalian University of Technology, Dalian 116024, China \\
$^2$Centre for Quantum Technologies and Department of Physics,
National University of Singapore, 3 Science Drive 2, Singapore
117543, Singapore\\
$^3$ Hefei National Laboratory for Physical Sciences at Microscale
and Department of Modern Physics, University of Science and
Technology of China, Hefei, Anhui 230026, China}

\date{\today}
\begin{abstract}
Identifying non-Markovianity with non-divisibility, we propose a
measure for non-Markovinity of  quantum process. Three examples are
presented to illustrate the non-Markovianity,  measure for
non-Markovianity is calculated and discussed. Comparison with other
measures of non-Markovianity is made. Our non-Markovianity measure
has the merit that no optimization procedure is required and it is
finite for any quantum process, which greatly enhances the practical
relevance of the proposed measure.
\end{abstract}
\maketitle

\section{introduction}

A quantum process is said to be Markovian  if  the future states of
the process depends only on the state of present time. In contrast,
dependence on past states should then be a characteristic feature of
non-Markovian processes. With the development of technology to
manipulate  quantum system, the quantum non-Markovian process has
attracted increasing attention in recent years
\cite{BreuerMC,Maniscalco,BreuerLin,Wolf,Piilo,Rebentrost}. On one
hand the inevitable interaction of a quantum system with its
environment leads to dissipation of energy and loss of quantum
coherence, on the other hand the quantum system may temporarily
regain some of the previously lost energy and/or information due to
non-Markovian effects in the  dynamics. This motivates the study on
the non-Markovianity and a  measure for the degree of
non-Markovianity is indeed needed.

Several approaches are proposed to quantify non-Markovianity,
including the measure   based on the increase of trace distance
\cite{Breuer-td}, the measure by quantifying the increase of
entanglement shared between the system and an isolated ancilla and
the divisibility of the dynamical map\cite{Rivas}, the measure based
on the decay rate of master equation itself \cite{Andrsson}, and the
measure through the Fisher information flow\cite{Lu}. Although
several approaches to quantifying the non-Markovianity are proposed,
the definition of non-Markovianity still remains elusive and, in
some sense, controversial \cite{mazzola10}.

It has been proven that all divisible dynamical maps are Markovian,
this divisibility property holds for a  larger class of quantum
processes than those described by the Lindblad master equation, for
example,  the  time-local master equation  with positive decay
rates. This indicates that the divisibility may be a good starting
point to quantify  non-Markovianity. In this paper, we will propose
a measure for non-Markovianity based on the divisibility of
dynamical maps, three dynamical maps are presented and the
corresponding non-Markovian measures are calculated. These results
suggest that the measure can capture   the feature  of non-Markovian
dynamics, and provide an easy way to calculate the non-Markovianity.

The paper is organized as follows: In Sec.\rm{II},  we discuss the
non-divisibility of dynamical map and the non-Markovianity of this
map. In Sec.\rm{III}, we introduce a measure for non-Markovianity.
Three examples are presented and discussed in Sec.\rm{IV}.
Sec.\rm{V} summarizes our results.

\section{non-divisibility and non-Markovianity}
In quantum  Markovian process,   the future state of the quantum
system  depends only on the state of present time. However, writing
this statement in  a precise mathematical representation is not an
easy task. Instead, we use the following description for quantum
Markovian process. A quantum evolution is Markovian if it is an
element of any one-parameter continuous completely positive
semigroup \cite{Alicki}. The quantum evolution governed by the
master equation
\begin{eqnarray}
\frac{d\rho}{dt}=\mathcal{L}\rho, \label{eqn:ME}
\end{eqnarray}
is an example,  where $\mathcal{L}$ is a time-independent generator
of the well-known Lindblad form,
\begin{eqnarray}
\mathcal{L}\rho&=&-i[H,\rho]\nonumber\\
&+&\sum_\alpha\gamma_a(V_\alpha\rho
{V_\alpha}^\dag-\frac{1}{2}{V_\alpha}^\dag V_\alpha \rho-
\frac{1}{2}\rho{V_\alpha}^\dag V_\alpha ) \label{eqn:MGen}
\end{eqnarray}
with $\gamma_{\alpha}\geq0$. This generator leads to
completely positive trace-preserving maps $\Lambda(t)=e^{\mathcal{L}t}$
and it satisfies the composition law,
\begin{eqnarray}
\Lambda(t_1+t_2)=\Lambda(t_2)\Lambda(t_1).
\label{eqn:TIDComplaw}
\end{eqnarray}
If a dynamical map can be written in this decomposition with both
$\Lambda(t_2)$ and $\Lambda(t_1)$ being completely positive, the
dynamical  map is called divisible. This composition law can be
extended to a general  case, where the generator in Eq.
(\ref{eqn:MGen}) is  time-dependent, namely,
\begin{eqnarray}
\frac{d\rho}{dt}=\mathcal{L}(t)\rho
\label{eqn:TDMarME}
\end{eqnarray}
with
\begin{eqnarray}
\begin{split}
\mathcal{L}(t)\rho= -i[H(t),\rho]+\sum_\alpha\gamma_{\alpha}(t)
(V_\alpha(t)\rho {V_\alpha}^\dag(t)-\\\frac{1}{2}{V_\alpha}^\dag(t)
V_\alpha(t) \rho-\frac{1}{2}\rho{V_\alpha}^\dag(t) V_\alpha(t)), \ \
\ \ \ \ \ \ \label{eqn:TDMarGen}
\end{split}
\end{eqnarray}
where $\gamma_{\alpha}(t)\geq0$. This is known as time-dependent
markovian \cite{Alicki}. The solution to Eq. (\ref{eqn:TDMarME}) can
be written in terms of the two-parameter family of dynamical maps
$\Lambda(t_2,t_1)$ ($t_2\geq t_1\geq0$). The composition law
corresponding to Eq. (\ref{eqn:TIDComplaw}) becomes
\begin{eqnarray}
\Lambda(t_2,0)=\Lambda(t_2,t_1)\Lambda(t_1,0), \label{eqn:TDMarMaps}
\end{eqnarray}
and the map $\Lambda(t_2,t_1)$ can be written  as $
\Lambda(t_2,t_1)=Te^{\int_{t_1}^{t_2}\mathcal{L}(t') dt'}$, $T$ is
the chronological operator. The composition law (divisibility of the
map) implies that the dynamical map $\Lambda(t_2,t_1)$
\begin{eqnarray}
\Lambda(t_2,t_1)=T e^{\int_{t_1}^{t_2}\mathcal{L}(t')dt'}
\quad\quad(t_2\geq t_1\geq 0), \label{eqn:Marmap}
\end{eqnarray}
transforming a state at $t_1$ into a state at $t_2$  (for systems
governed by time independent master equation (\ref{eqn:ME}),
$\Lambda(t_2,t_1)=\Lambda(t_2-t_1,0)=\Lambda(t)$ ) must be
trace-preserving and completely positive, regardless of  which
dynamics it describes, e.g., it is from the time-dependent master
equation or the time-independent master equation. Note that the
starting time $t_1$ is not zero.

A  measure for non-Markovianity should quantify the  deviation of a
dynamical map from Markovian evolution. Noticing  that when a
dynamics  is non-Markovian, the dynamical  map  $\Lambda(t_2,t_1)$
may not  be completely positive, we may use the non-divisibility to
quantify the non-Markovianity. In fact,  this is the underlying
reason that the trace distance can increase \cite{Breuer-td}, and
the system gains entanglement  with an isolated ancilla
\cite{Rivas}.

It is worth stressing that there is no contradiction between the
requirement on  non-completely positivity  and that on   physics.
Consider a quantum evolution in a time interval $(0,t_2)$, we always
have $\Lambda(t_2,0)=\Lambda(t_2,t_1)\Lambda(t_1,0)$ due to the time
continuity. For $\Lambda(t_2,0)$ to be a dynamical map, it is
required that $\Lambda(t_2,0)$ must be completely positive, however,
$\Lambda(t_2,t_1)$ may not be completely positive. Therefore these
two-parameter maps in non-Markovian dynamics do not generate a
quantum dynamical semigroup.  Then one may wonder: does there exist
a $\Lambda(t_2,t_1)$ that it is not completely positive but
$\Lambda(t_2,0)$ does? The answer is yes. First,  a wide range of
non-Markovian process can be described by time-local master
equations via time-convolutionless projection operator
\cite{Chaturvedi,Shibata,Royer,Breuertcl,Breuer,Feynman}. Second, it
has been shown that any quantum dynamics described by memory kernel
master equation may be written  in a time-local form
\cite{Chruscinski}. Note that the decay rates in these time-local
master equation  are different from that in Eq.
(\ref{eqn:TDMarGen}), they can be negative. With this time-local
master equation, the dynamical map with non-zero starting time in
Eq.(\ref{eqn:Marmap})
 may violate the complete positivity due to the negative decay rates.
 This implies that the
non-complete positivity of the map  $\Lambda(t_2,t_1)$ is an
essential feature of non-Markovian process. The time-dependent decay
rate  may be negatively infinite at some  points of time,  where the
reviving of population or regaining of quantum coherence happen
\cite{Breuer,Breuer-td}. We call these points of time  singular
points $t_s$. When $t_1=t_s$ or $t_2=t_s$ , $\Lambda(t_2,t_1)$ may
not exist. However, we can use $\Lambda(t_2,t_1)$ in the limit that
$t_1\rightarrow t_s$ instead of $\Lambda(t_2,t_s)$. It is convenient
to discuss $\Lambda(t_2,t_1)$ with a specific  time-local master
equation, but this is not necessary.

\section{Measure for non-Markovianity}

To construct a measure for non-Markovianity, we resort to  the
Choi-Jamio{\l}kowski isomorphism \cite{Choi,Jamio}, it asserts  that
a linear map $\Lambda$ :$M_d\rightarrow M_d$ is isomorphic to the
Choi matrix,
\begin{eqnarray}
C_{\Lambda}=\sum_{i,j=1}^d\ket{i}\bra{j}\otimes\Lambda(\ket{i}\bra{j}),
\end{eqnarray}
where $\ket{i}$ are orthogonal bases. An  familiar form of the Choi
matrix is
\begin{eqnarray}
\rho_{\Lambda}=(\Lambda\otimes \texttt{I})\ket{\phi}\bra{\phi},
\label{eqn:rhol}
\end{eqnarray}
where $\ket{\phi}$ is the maximally entangled  state
$\ket{\phi}=\frac{1}{\sqrt{d}}\sum_{i=1}^d\ket{i}\otimes\ket{i}$,
$\texttt{I}$ is an identity map acting on the ancilla, and
$\rho_{\Lambda}$ is the normalized $C_{\Lambda}$. It turns out that
$\Lambda$ is completely positive if and only if $\rho_{\Lambda}$ is
positive semidefinite. In   other words, the sufficient and
necessary condition of non-complete positivity is the negativity of
$\rho_{\Lambda}$. Hence, the sum of negative eigenvalues of
$\rho_{\Lambda}$ can be taken as a measure for the non-complete
positivity of the dynamical map. However, the summation may
sometimes be an infinite value in some models due to singular decay
rates, this suggests to use a normalized quantity
\begin{eqnarray}
\it{Ncp}=\arctan(-\sum\lambda_k) \label{ncp}
\end{eqnarray}
as a measure for non-complete positivity of the map $\Lambda,$ where
$\lambda_k$ is the $k-{th}$ negative eigenvalue of $\rho_{\Lambda}$.
Clearly if  $\rho_{\Lambda}\geq 0$, $\it{Ncp}$ $=0$. Then we have
$0\leq \it{Ncp} \leq \frac{\pi}{2}$. Based on the complete
positivity property of the map $\Lambda(t_1,t_2)$, a measure of
non-Markovianity has been proposed \cite{Rivas}. This measure is
different from ours in that we use the an averaged negativity of the
map $\Lambda(t_1,t_2)$ to quantify the non-Markovianity.  Moreover,
insightful examples are presented to shed light on the
non-Markovianity measure.

To calculate Eq. (\ref{ncp}) with a given time-local master
equation, the exact form of $\Lambda(t_1,t_2)$ is not necessary.
What we need is  to extend the time-local master equation from one
system to two systems, taking an isolated ancilla attached to the
system. The Hilbert space is extended from  $H_d$ to $H_d\otimes
H_d$ accordingly. All operators, say $\hat{O}$,  are replaced by
$\hat{O}\otimes \texttt{I}.$  By this extension, we get a new master
equation, which describes the system and the ancilla. The system
evolves in the same manner as before, while the anciila is isolated
from both the system and environment. The master equation can be
solved starting from $\ket{\phi}$ at time $t_1$ and the state at
time $t_2$ ($t_2>t_1$) can be obtained.

We aim at finding a measure $NM$ for non-Markovianity which captures
the feature of  non-complete positivity of all possible
$\Lambda(t_2,t_1)$. Note that $\it{Ncp}$  is a function of $t_1$ and
$t_2$.  Let $S$ count the number  of $\it{Ncp}$ in all time interval
with $\it{Ncp}$ $>0$, i.e.
\begin{eqnarray}
S=\int_0^\infty\!dt_1\int_{t_1}^\infty\!dt_2\;c(t_2,t_1), \quad
c=\left\{\begin{array}{c} 1,\  if\  \it{Ncp}>0 \\ 0, \ if \ \it{Ncp}=0.\\
\end{array}
\right.
\label{eqn:area}
\end{eqnarray}
If $S=0$, i.e., all $\Lambda(t_2,t_1)$ (for any $t_1$ and $t_2$, as
long as $t_2>t_1$) are completely positive, the non-Markovinity $NM$
should be  defined  to be zero. If $S>0$, we define
\begin{eqnarray}
NM=\lim_{T\rightarrow+\infty}
\frac{\int_0^T\!dt_1\int_{t_1}^T\!dt_2\  \it{Ncp}\ (t_2,t_1)}
{\int_0^T\!dt_1\int_{t_1}^T\!dt_2\;c(t_2,t_1)} \label{eqn:define}
\end{eqnarray}
as a measure  of non-markovianity. This can be understood  as an
averaged non-complete positivity of all the non-completely positive
maps in interval $t=(0,+\infty)$. Therefore, from $0\leq
\it{Ncp}\leq\frac{\pi}{2}$, we have $0\leq NM\leq\frac{\pi}{2}$ .

The upper limit of the integral  $T$ in the definition  Eq.
(\ref{eqn:define}) is taken to be infinite. However, the
distribution of $\it{Ncp}$ on $(t_1, t_2)$ plan  is often periodic
or is limited in a small  area, this suggests that the integration
can be taken merely in one period of time  or taken in a finite
area. When $\it{Ncp}$ is neither periodic nor limited in a small
region, the upper limit $T$ in Eq. (\ref{eqn:define}) should be
large enough to get a convergent $NM$. The definition can be written
into a simple form,
\begin{eqnarray}
NM=\textrm{E}(\it{Ncp}(\Lambda_N)), \label{eqn:exception}
\end{eqnarray}
where $\Lambda_N$ represents all the non-completely  positive maps
and $\textrm{E}$ means expectation  value. Therefore, $NM$ can be
numerically calculated by averaging a large number of non-completely
positive maps with equal weight  (or randomly) in a {\it
representative area} as discussed above.  The representative area
means one period in time or a limited region where $\it{Ncp}>0.$

To illustrate the measure of non-Markovianity, we present three
examples in the next section. We work in the interaction picture for
simplicity to calculate the measure, since  unitary transformation
does not change the eigenvalues of  $\rho_{\Lambda}$ as well as
$\it{Ncp}$ of $\Lambda(t_2,t_1)$, hence the non-Markovianity measure
under unitary transformation  remains unchanged.

\section{examples}
\subsection{damping J-C model}

The first example is  a two-level system coupling to  a reservoir at
zero temperature. The reservoir consists of infinite number of
harmonic oscillators that  is also referred in the literature as the
spin-boson model. This model  is exactly solvable \cite{Breuer}. The
Hamiltonian for such a  system reads,
\begin{eqnarray}
H=H_0+H_I
\end{eqnarray}
\begin{eqnarray}
\textrm{with} \qquad H_0=\hbar\omega_0\sigma_+\sigma_-
 +\sum_k\hbar\omega_k b_k^{\dag}b_k\qquad\qquad,
\\H_I=\sigma_+B+\sigma_-B^\dag \qquad\qquad\qquad,
\end{eqnarray}
where $B=\sum_k g_k b_k$. The Rabi frequency  of the two-level
system and the frequency for the $k-th$ harmonic oscillator  are
denoted by $\omega_0$ and $\omega_k,$ respectively. $b^\dag_k$ and
$b_k$ are the creation and annihilation operators of  $k-th$
oscillator, which  couples to the system with  coupling constant
$g_k$.

Assuming  the system  and the reservoir are initially uncorrelated,
we can obtain a   time-dependent master equation in the interaction
picture,
\begin{eqnarray}
\dot{\rho}&=&-i\frac{s(t)}{2}[\sigma_+\sigma_-,\rho]\nonumber\\
&+&\gamma(t)(\sigma^-\rho\sigma^+-\frac{1}{2}\sigma^+
\sigma^-\rho-\frac{1}{2}\rho\sigma^+\sigma^-), \label{eqn:exactnm}
\end{eqnarray}
where $s(t)=-2\mathrm{Im}[\frac{\dot{c}(t)}{c_0(t)}]$  and
$\gamma(t)=-2\mathrm{Re}[\frac{\dot{c}(t)}{c_0(t)}]$. $\Omega(t)$
plays the role  of Lamb shift and $\gamma(t)$ is the decay rate.
Both $\Omega(t)$ and $\gamma(t)$ are  time-dependent. $c(t)$ is
determined by $\dot{c}(t)=-\int_0^t f(t-\tau)c(\tau)d(\tau)$, where
$f(t-\tau)=\int d\omega J(\omega) exp(i(\omega_0-\omega)(t-\tau))$
is the environmental correlation function. In the derivation of the
master equation, the reservoir is assumed in its vacuum at $t=0$.

Consider the following spectral density,
\begin{eqnarray}
J(\omega)=\frac{1}{\pi}\frac{\gamma_0\lambda^2}
{(\omega_0-\omega)^2+\lambda^2}, \label{sd}
\end{eqnarray}
where $\gamma_0$ represents the coupling constant between the system
and reservoir, $\lambda$ defines the spectral width of  the coupling
at the resonance point $\omega_0$. For the spectral density
(\ref{sd}), we have  $s(t)=0$, $c(t)=c_0 e^{-\lambda
t/2}[\cosh(\frac{dt}{2})+\frac{\lambda}{d}\sinh(\frac{dt}{2})]$, and
$\gamma(t)=\frac{2\gamma_0\lambda
\sinh(dt/2)}{d\cosh(dt/2)+\lambda\sinh(dt/2)}$ with
$d=\sqrt{\lambda^2-2\gamma_0\lambda}$. Note that Eq.
(\ref{eqn:exactnm}) is derived  without any approximations, hence it
is non-Markovian and it exactly describes the dynamics of the open
system.

It is well-known  that $\lambda$ characterizes the  correlation time
$\tau_R$ of the reservoir through $\tau_R=\lambda^{-1}$. The time
scale $\tau_S$ on which the state of the system changes is given by
$\tau_S=\gamma_0^{-1}$. So the degree of non-Markovianity should be
relevant to the rate $R=\tau_R/\tau_S$. Namely, when $R$ is very
small, the evolution is  Markovian,  when $\tau_R$ is comparable
with $\tau_S$, the memory effect of reservoir should  be taken into
account, the dynamics of the open system is then  non-Markovian.

Let us  first analyze the non-Markovianity of the dynamics by
examining  $\gamma(t)$. For $R<\frac{1}{2}$, $\gamma(t)$ is always
positive and it is a monotonically increasing function of time,  all
$\Lambda(t_2,t_1)$ are completely positive, hence the dynamics is
Markovian. When $R>\frac{1}{2}$, $\gamma(t)$ is a periodic function
of time,  it takes negative values sometimes. In particular,
$\gamma(t)$ has discrete singular points where the upper level gains
population, this is a typical feature of non-Markovianity.

Now we see if our measure can capture all these features of
non-Markonianity. In order to apply our measure, we have to extend
the time-local master equation to the compound system, i.e.,  the
operators $\sigma_\pm$ in Eq. (\ref{eqn:exactnm}) is replaced by
$\sigma_\pm\otimes \texttt{I}$ with $\texttt{I}$ being the ancilla's
$2\times 2$ identity operator. For a given time interval
$(t_1,t_2)$, a straightforward calculation yields
\begin{equation}
\it{Ncp}(t_1,t_2)=\left\{\begin{array}{c}
\arctan(\frac{1}{2}(\frac{c(t_2)^2}{c(t_1)^2}-1)),
\quad\quad\frac{c(t_2)^2}{c(t_1)^2}>1\\
\quad\quad\quad\quad 0, \quad\quad\quad\quad\quad\quad\ \
\frac{c(t_2)^2}{c(t_1)^2}\leq 1,\\
\end{array}
\right.
 \end{equation}
where $c(t)$ was defined below Eq. (\ref{eqn:exactnm}) and $t_2\geq
t_1\geq 0$.
\begin{figure}
\includegraphics*[width=6.5cm]{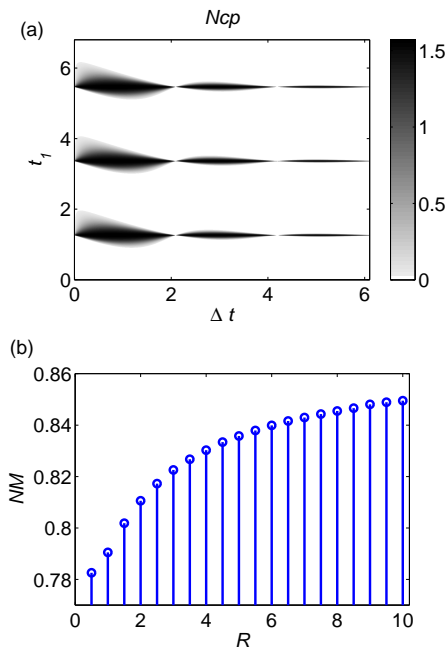}
\caption{$\it{Ncp}$ and non-Markovianity in the damping J-C model.
(a) $\it{Ncp}$ versus  $t_1$  and $\Delta t$ (in units of
$1/\lambda$) at $R=5$. Three oscillations  are shown. (b) $NM$ as a
function of $R$. Clearly,  $NM$ monotonically increases with $R$.}
\label{FIG:jc}
\end{figure}

For a typical non-markovian case ($R=5$), we plot $\it{Ncp}$ as a
function of $t_1$ and $\Delta t$  in Fig.\ref{FIG:jc}(a) ($\Delta
t=t_2-t_1$, since $t_2\geq t_1$, we use $\Delta t$ instead of $t_2$
for convenience). This plot shows the non-zero area and its value of
$\it{Ncp}$ versus $t_1$ and $t_1-t_2$. As $\it{Ncp}$ is a periodic
function of $t_1$, and the area where $\it{Ncp}$ $>0$ decays very
fast with $\Delta t.$ $NM$ can be given by averaging all
 $\it{Ncp}$ in one period, it yields $NM=0.835.$

As expected, when $R<0.5$, $NM=0$, and when $R>0.5$ the
non-markovianity is finite. We plot the measure of non-Markovianity
with different $R$ in Fig.\ref{FIG:jc}(b). Here $R$ is chosen from
$0.5$ to $10$. Note that when $R=0.5$, $\gamma(t)$ does not exist
due to the zero denominator. Our result for $R=0.5$ is obtained at
$R=0.5+\varepsilon$ with $\varepsilon$ an infinitesimal positive
number. Intuitively, the larger $R$ is, the stronger the
non-Markovianity. The results in Fig.\ref{FIG:jc}(b) demonstrate
that this is  indeed the case.

\subsection{J-C model with detuning}
\begin{figure}
\includegraphics*[width=6.5cm]{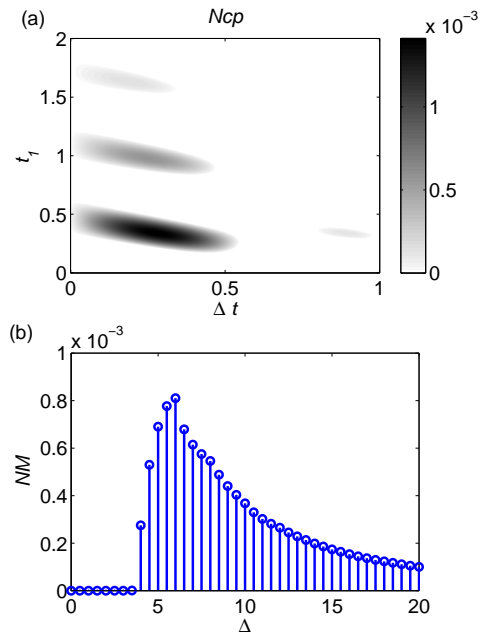}
\caption{ $\it{Ncp}$ and non-Markovianity in the detuning J-C model.
(a)Counter plot of  $\it{Ncp}$ as a function of  $t_1$ and $\Delta
t$ (in units of $1/\lambda$), with $\Delta=10$ and $\gamma_0=0.3$
(in units of $\lambda$). (b) $NM$ as a function of $\Delta$. Other
parameters chosen are the same as in (a). The maximum non-Markianity
arrives  at about $\Delta=6$.} \label{FIG:dt}
\end{figure}

The second example is similar to the first one, but here we consider
the system in a cavity whose  center frequency is detuned from the
system  Rabi frequency $\omega_0$. The  dynamics in the interaction
picture is governed  by Eq. (\ref{eqn:exactnm}), but $s(t)$ and
$\gamma(t)$ are  determined by the Lorenz spectral density
\begin{eqnarray}
J(\omega)=\frac{1}{2\pi}\frac{\gamma_0\lambda^2/2}
{(\omega_0-\Delta-\omega)^2+\lambda^2},
\end{eqnarray}
where $\Delta$ denotes the  detuning.

We extend Eq. (\ref{eqn:exactnm}) to a  compound system by
introducing  an ancilla as we did in the first example, then we
calculate $\it{Ncp}(t_1,t_2)$ numerically. We plot $\it{Ncp}$ as a
function of $t_1$ and $\Delta t$  in Fig.\ref{FIG:dt}(a) for a
typical non-Markovian case. We see that $\it{Ncp}(t_1,t_2)$  is
finite in contrast to infinite values  in the same region  for the
first example. On the other hand, $\it{Ncp}$ of all
$\Lambda(t_2,t_1)$ (with different $t_1$ and $t_2$)  are far less
than $\frac{\pi}{2},$ indicating weaker non-Markovianity  in
comparison  with the first example. Finally, from Eq.
(\ref{eqn:define}), or Eq. (\ref{eqn:exception}) we have
$NM=3.66\times10^{-4}$ in this case.

Now we discuss the dependence of non-Markovianity  on the  detuning.
We plot $NM$ in Fig.\ref{FIG:dt}(b) with different $\Delta$.  We
find  that $NM$ appears non-zero  at about $\Delta=4$, it first
increases then decreases with $\Delta$. This result is similar to
that in \cite{Breuer-td}, where the non-Markovianity is  measured by
the decreases of trace distance.

\subsection{a two-level system coupling to  a  finite spin bath}
In the third example, we consider a central spin-$\frac 12$ coupling
to a bath of $N$ spin-$\frac 1 2$s. The interaction Hamiltonian is,
\begin{equation}
H=\sum_{k=1}^NA_k\sigma_z\sigma_z^{k}
\end{equation}
where $A_k=A/\sqrt{N}$ represents the coupling constants. The
non-Markovianity of the central spin in this model is discussed in
\cite{BreuerMC}. Assume the initial state of the whole  system is
$\rho_s(0)\otimes (\frac{1}{2^N}I)$, i.e.,  all  spins in the
reservoir  are in  a maximal mixed state. The density matrix of the
central spin at time $t$ takes,
\begin{equation}
\rho(t)=
\left(
\begin{array}{cc}
\rho_{11} & \rho_{12}\cos^N(\frac{2At}{\sqrt{N}})\\
\rho_{21}\cos^N(\frac{2At}{\sqrt{N}}) & \rho_{22}\\
\end{array}
\right).\label{denc}
\end{equation}
In terms of dynamical map, the dynamics can be represented as,
$\Lambda(t,0)\rho=\frac 1 2 (1-\cos^N(\frac{2At}{\sqrt{N}}))
\sigma_z\rho\sigma_z+\frac{1}{2}(1+\cos^N(\frac{2At}{\sqrt{N}}))\rho.$
This is equivalent to  the following master equation,
\begin{equation}
\dot{\rho}=\gamma(t)\mathcal{L}(\rho), \label{eqn:tan}
\end{equation}
where $\mathcal{L}(\rho)=\sigma_z\rho\sigma_z-\rho$,   and the
time-dependent $\gamma(t)=\frac{N}{2}\tan(\frac{2At}{\sqrt{N}})$.
This example is discussed in several papers as a classical example
to quantify non-Markovianity,  and the  non-Markovianity
 is infinite\cite{Breuer-td,Rivas}. By our definition of
non-Markovianity, it is finite. This allows us to establish a
relation between non-Markovianity $NM$ and the number of spin $N$ in
the reservoir.

\begin{figure}
\includegraphics*[width=6.5cm]{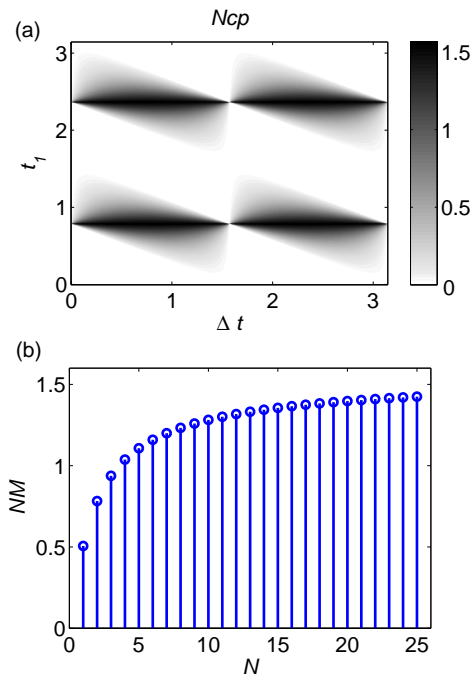}
\caption{$\it{Ncp}$ and non-Markovianity in  the N-spin bath model.
(a) $\it{Ncp}$ as a function of $t_1$ and $\Delta t$ for $N=1$.  $t$
is in units of $\frac{1}{A}$. It shows that $\it{Ncp}$ is a periodic
function of  $t_1$ and $t_2$, two periods are  shown here. (b) $NM$
with different $N$. Non-Markovianity increases with $N$ and tends to
$\frac{\pi}{2}$ when the limit $N\rightarrow+\infty$ is taken after
$T\rightarrow \infty$.} \label{FIG:tan}
\end{figure}

It is easy to calculate  $\rho_{\Lambda}$ (defined in
Eq.(\ref{eqn:rhol})) by Eq. (\ref{eqn:tan}),
\begin{equation}
(\Lambda(t_2,t_1)\otimes \texttt{I})\ket{\psi}\bra{\psi}= \left(
\begin{array}{ccccc}
  0.5 & 0 & 0 & 0.5k \\
  0 & 0 & 0 & 0 \\
  0 & 0 & 0 & 0 \\
  0.5k & 0 & 0 & 0.5 \\
\end{array}
\right)\label{denc1}
\end{equation}
with $k=\frac{\cos^N(\frac{2At_2}{\sqrt{N}})}
{\cos^N(\frac{2At_1}{\sqrt{N}})}\ \  (t_2\geq t_1\geq 0)$, then we
have
\begin{equation}
\it{Ncp}(t_1,t_2)=
\left\{
\begin{array}{c}
\arctan(\frac{1}{2}(|k|-1)),\quad\quad\quad\  (|k|>1)\\
 \quad\quad\quad 0, \quad\quad\quad\quad\quad(-1\leq k\leq1).
\end{array}
\right.
\end{equation}

The non-completely positivity of  the map $\Lambda(t_2,t_1)$ can be
examined  by plotting  $\it{Ncp}$ in ($t_1, t_2$) plan.
Fig.\ref{FIG:tan}(a) shows this result taking only one spin as the
environment (for $N>1$, the results is similar). We find that
$\it{Ncp}$ is a periodic function of  both $t_1$ and $\Delta t$.
When $t_1=(2n+1)\frac{\pi}{4}$ and $\Delta t\neq n\frac{\pi}{2}$
$(n=0,1,2\cdots)$,  $\it{Ncp}=\frac{\pi}{2}$ reaches its maximum.
The  non-Markovianity measure in this case is $NM=0.505.$

Now we try to find the relation between non-Markivianity  and the
number of  spins $N$ in the environment, the result is plotted in
Fig.\ref{FIG:tan}(b). We find that  $NM$ increases with the spin
number $N$. For a very large $N$, $NM$ arrives at its maximum.
However, the coupling strength $A_k=A/\sqrt{N}$  is very small in
large $N$ limit, this implies that the characteristic time of the
central spin tends to be infinitely long in this limit, i.e.,
non-Markovian dynamics can be observed only on a long time scale. If
we are interested in a limited time interval $(0,t)$, all dynamical
maps $\Lambda(t_1,t_2)$ are completely positive, there is no
non-Markovian effect. This can be understood as follows.  With a
fixed $T$, one always can choose $N$ that $\frac{2A T}{\sqrt N}$ is
close to zero. The off-diagonal element of the density matrix
Eq.(\ref{denc1}) then takes $\rho_{12} e^{-2A^2(t_2^2-t_1^2)}$,
indicating that the density matrix in Eq.(\ref{denc1}) describes a
typical Markovian process.

\section{summary}
We have  presented  a measure for non-markovianity based on the
divisibility of dynamical map. This measure has the advantage that
it is easy to calculated and no optimization is required. Three
examples are illustrated, which show that the measure can nicely 
manifest the non-Markovianity. We also compare our measure with
others in the literature and find  that it almost agrees with the
trace-distance based measure.

This work is supported by NSF of China under  grant Nos 61078011 and
10935010, as well as the National Research Foundation and Ministry
of Education, Singapore under academic research grant No. WBS:
R-710-000-008-271.

\end{document}